\documentclass[12pt]{article}
\usepackage{graphicx}
\usepackage{amsmath}
\usepackage{amssymb}

\leftmargin=--0.3in   \topmargin=-0.20in
\newcommand\pubnumber{}
\newcommand\pubdate{}

\def\orsay{Laboratoire de Physique Th\'eorique\\
CNRS/Universit\'e Paris-Sud, B\^at 210, F-91405 Orsay Cedex, FRANCE}

\def\Title#1{\begin{center} {\Large #1 } \end{center}}
\def\Author#1{\begin{center}{ \sc #1} \end{center}}
\def\Address#1{\begin{center}{ \it #1} \end{center}}

\newcommand\pubblock{\rightline{\begin{tabular}{l} \pubnumber\\
         \pubdate  \end{tabular}}}
\newenvironment{Abstract}{\begin{quotation}  }{\end{quotation}}
\newenvironment{Presented}{\begin{quotation} \begin{center} 
             PRESENTED AT\end{center}\bigskip 
      \begin{center}\begin{large}}{\end{large}\end{center} \end{quotation}}

\def\beq{\begin{equation}}
\def\eeq#1{\label{#1}\end{equation}}
\def\eeqn{\end{equation}}

\def\beqa{\begin{eqnarray}}
\def\eeqa#1{\label{#1}\end{eqnarray}}
\def\eeqan{\end{eqnarray}}

\let\bar=\overbar

\def\Dslash{\not{\hbox{\kern-4pt $D$}}}
\def\dslash{\not{\hbox{\kern-2pt $\del$}}}

\def\msb{{\bar{\ssstyle M \kern -1pt S}}}

\def\rgl{\rangle}
\def\lgl{\langle}
\def\gc{\gamma^5}
\def\taud{\tau_{1/2}}
\def\taut{\tau_{3/2}}
\def\m{\mu}
\def\elematrice#1#2#3{\lgl#1|#2|#3\rgl}

\newcommand{\rb}[1]{\raisebox{1.5ex}[0pt]{#1}}

\begin{document}
\begin{titlepage}
\pubblock

\vfill
\Title{$B \to D^{**}$ -- puzzle 1/2 vs 3/2}
\vfill
\Author{Beno\^\i t Blossier} 
\Address{\orsay}
\vfill
\begin{Abstract}
Understanding the composition of final states in $B \to X_c l \nu$ could help to get a feedback on the persisting disagreement between exclusive and inclusive determinations of $V_{cb}$. In particular the series of orbital excitations $D^{**}$ and radial excitations ($D'$, $D^*\,'$) has received a lot of attention; a misinterpretation as a scalar state of the ($D' \to D \pi$) spectrum tail could have induced an experimental overestimate of the broad states contribution to the total $B \to X_c l \nu$ width with respect to theoretical expectations, all of them made however in the infinite mass limit: it is the so-called 1/2 vs 3/2 puzzle. We describe first attempts to measure on the lattice form factors of $B \to D^{**} l \nu$ at realistic quark masses. Cleaner processes, like hadronic decays $B \to D^{**} \pi$ and semileptonic decays $B_s \to D_s^{**} l \nu$ in the strange sector have recently been examined by phenomenologists, putting new interesting ideas on those issues with, again, the need of lattice inputs.
\end{Abstract}
\vfill
\begin{Presented}
the 8th International Workshop on the CKM Unitarity Triangle (CKM 2014), Vienna, Austria, September 8-12, 2014
\end{Presented}
\vfill
\end{titlepage}
\def\thefootnote{\fnsymbol{footnote}}
\setcounter{footnote}{0}

\section{Introduction}

Understanding the long-distance dynamics of QCD is crucial in the control of the theoretical systematics on low-energy processes that are investigated at LHCb and, in the next years, at Super Belle, to detect indirect effects of New Physics. It is particularly relevant for processes involving excited states, that occur often in experiments. With that respect beauty and charmed mesons represent a very rich sector. An intriguing question concerns the origin of the $\sim 3 \sigma$ discrepancy between $|V_{cb}|^{\rm excl}$ and $|V_{cb}|^{\rm incl}$ \cite{vckmfit}: expressed differently, it is welcome to know more about the composition of the final hadronic state $X_c$ in the semileptonic decay $B \to X_c l \nu$. We sketch in Table \ref{tabspectrum} the low-lying spectrum of $D$ mesons.
\begin{table}
\begin{center}
\begin{tabular}{|c|c|c|c|c|c|}
\cline{3-6}
\multicolumn{1}{c}{}&\multicolumn{1}{l|}{}&\rule[1mm]{0mm}{3mm}Mass (MeV)&\rule[1mm]{0mm}{3mm}Width (MeV)
&\rule[1mm]{0mm}{3mm}$j_l^P$&\rule[1mm]{0mm}{3mm}$J^P$\\
\hline
&\rule[1mm]{0mm}{3mm}$D^\pm$&\rule[1mm]{0mm}{3mm}1869$\pm$0.5&-&&\rule[1mm]{0mm}{3mm}$0^-$\\
\rb{$S$: $D^{(*)}$}&$D^{*\pm}$&2010$\pm$0.4&96$\pm$25&\rb{$\frac{1}{2}^{-}$}&$1^-$\\
\hline
&\rule[1mm]{0mm}{3mm}$D^*_0$&\rule[1mm]{0mm}{3mm}2352$\pm$ 50&\rule[1mm]{0mm}{3mm}261 $\pm$ 50
&&\rule[1mm]{0mm}{3mm}$0^+$\\
&$D^*_1$&2427$\pm$ 26 $\pm$ 25& $384^{+107}_{-75}\pm 74$&\rb{$\frac{1}{2}^{+}$}&$1^+$\\
\cline{2-6}
\rb{$P$: $D^{**}$}&$D_1$&2421.8 $\pm$ 1.3&$20.8^{+3.3}_{-2.8}$&&$1^+$\\
&$D^*_2$&2461.1$\pm$ 1.6&32$\pm$ 4&\rb{$\frac{3}{2}^{+}$}&$2^+$\\
\hline
\end{tabular}
\end{center}
\caption{\label{tabspectrum} Low-lying spectrum in the $D$ sector; it is convenient to decompose the total orbital momentum as $J=\frac{1}{2} \oplus j_l$, where $j_l$ is the orbital momentum of the light degrees of freedom.}
\end{table}
The D states of the $j^P_l=\frac{1}{2}^+$ doublet are broad while those of the $j^P_l=\frac{3}{2}^+$ doublet 
are narrow: indeed, the main decay channels are the non leptonic transitions  $D^{**} \to D^{(*)}\pi$.
Parity conservation implies that the pion has an even angular momentum $\ell$ with respect to
$D^{(*)}$. Orbital momentum conservation implies that $\ell = 0$ or $2$. That's why $D^*_0$ and $D^*_1$ decay 
with a pion in the $S$ wave and $D^*_2$ decays with the pion in the $D$ wave. The decay $D_1 \to D^* \pi$ occurs with the pion in the $S$ or $D$ waves; however, thanks to Heavy Quark Symmetry, the latter is favored.  
Therefore, decays of the $j^P_l=\frac{3}{2}^+$ doublet are suppressed compared to decays of the $\frac{1}{2}^+$ doublet.
But $X_c$ could be made of radial excitations as well: the Babar Collaboration claimed to have isolated a bench of new $D$ states \cite{babard}. Among them, a structure in the $D^*\pi$ distribution is interpreted as $D(2550)\equiv D'$.
After a fit, experimentalists obtain $m(D')=2539(8)$ MeV and $\Gamma(D')=130(18)$ MeV.
A question raised about the correctness of this interpretation because, in theory, quark models predict approximately the same $D'$ mass (2.58 GeV) but a quite smaller width (70 MeV) \cite{quarkmodels}. However a well known caveat is that excited states properties are very sensitive to the position of the wave functions nodes, themselves depending strongly on the quark model. We collect in Table \ref{tabSL} the branching ratios of the $B \to X_c$ semileptonic decays.
\begin{table}
\begin{center}
\begin{tabular}{ccc}
\begin{tabular}{|l|}
\hline
${\cal B}(B_d \to X_c l \nu)=(10.09 \pm 0.22) \%$\\
${\cal B}(B_d \to [\mbox{non} - D^{(*)}] l \nu) = 2.86 \pm 0.25)\%$\\
${\cal B}(B_d \to D^{**}_{\rm narrow} l \nu) = (0.87 \pm 0.06) \%$\\
${\cal B}(B_d \to D^{(*)} \pi l \nu) = (1.43 \pm 0.08) \%$\\
${\cal B}(B_d \to [D \pi]_{\rm broad} l \nu)=(0.42 \pm 0.06) \%$\\
${\cal B}(B_d \to [D^* \pi]_{\rm broad} l \nu)=(0.33 \pm 0.07) \%$\\
\hline
\end{tabular}
&
\begin{tabular}{|c|c|}
\hline
$B_d \to D^{**} e \nu$&${\cal B}_{\rm exp}/ {\cal B}_{\rm th}$\\
\hline
$D^*_2$&0.5\\
$D_1$&1\\
$D^*_1$&[0,\,5]\\
$D^*_0$&$6 \pm 1$\\
\hline
\end{tabular}
&
\begin{tabular}{|c|c|}
\hline
$B_d \to D^{**} \pi$&${\cal B}_{\rm exp}/ {\cal B}_{\rm th}$\\
\hline
$D^*_2$&$\sim$ 0.5\\
$D_1$&[0.5,\,1]\\
$D^*_1$&no result\\
$D^*_0$&[0.2,\,2.6]\\
\hline
\end{tabular}
\end{tabular}
\end{center}
\caption{\label{tabSL}Branching ratio of $B \to X_c l \nu$ (left panel); comparison between theory
and experiment for the different $B \to D^{**} l \nu$ channels (center panel); comparison between theory
and experiment for the different $B \to D^{**} \pi$ channels (right panel).}
\end{table}
We are interested by $\sim 25\%$ of the total width $\Gamma (B \to X_c l \nu)$: 1/3 of it comes from
the channel $B \to D^{**}_{\rm narrow}$. Studying the channel $B \to D' l \nu$, assuming it is quite large \cite{bernlochner} and using the fact that $\Gamma (D' \to D_{1/2} \pi) \gg \Gamma(D' \to D_{3/2} \pi)$, one concludes that an excess of $B \to (D_{1/2}\pi) l \nu$ events could be observed with respect to their $B \to (D_{3/2} \pi) l \nu$ counterparts. A question is then whether such a potentially large $B \to D' l \nu$ width could explain the "1/2 vs. 3/2" puzzle: $[\Gamma (B \to D_{1/2} l \nu) \simeq \Gamma (B \to D_{3/2} l \nu)]^{\rm exp}$ while 
$[\Gamma (B \to D_{1/2} l \nu) \ll \Gamma (B \to D_{3/2} l \nu)]^{\rm theory}$ \cite{puzzle}. A kinematical
factor explains partly this suppression: $\frac{d\Gamma^{B \to D_{1/2}}}{d \Gamma^{B \to D_{3/2}}}=\frac{2}{(w+1)^2} \left(\frac{\tau_{1/2}(w)}{\tau_{3/2}(w)}\right)^2$. A detailed comparison between theory and experiment is
made in the center panel of Table \ref{tabSL}. The main tension is for
$B \to D^*_0 l \nu$. On the experimental side, there are issues about  
identifying the $D^*_0$ state and the disagreement in ${\cal B}(B \to D^*_1 l \nu)$ between 
Belle (no events) and BaBar (claim of a signal). On the theory side, the limitation is that the predictions are made 
essentially in the infinite mass limit, including lattice QCD calculations of
Isgur-Wise functions $\tau_{1/2}$ and $\tau_{3/2}$. 

\section{$B \to D^{**} l \nu$ and lattice QCD}

\subsection{Infinite mass limit}

In the Heavy Quark Effective Theory framework, with the trace formalism, the transitions between two heavy-light mesons $H^{j_l,J}_v$ and $H^{j'_l,J'}_{v'}$ are expressed in terms of universal form factors, the Isgur-Wise functions $\Xi(w\equiv v\cdot v')$, where $v$ is the velocity of the meson. Their number is limited thanks to Heavy Quark Symmetry:
$\xi(w)$ parameterizes the elastic transition $H^{\frac{1}{2}^-}_v \to
H^{\frac{1}{2}^-}_{v'}$ and is normalised at zero recoil: $\xi(1)=1$. One has also
$\elematrice{H^{0^+}_{v'}}{\bar{h}_{v'}\gamma^\mu \gc h_v}{H^{0^-}_v} = \taud(\mu,w) (v-v')^\mu$ and
$\elematrice{H^{2^+}_{v'}}{\bar{h}_{v'}\gamma^\mu \gc h_v}{H^{0^-}_v}
=\sqrt{3}\,\taut(\mu,w)[(w+1)\epsilon^{*\mu \alpha} v_{\alpha}
-\epsilon^*_{\alpha\beta}v^\alpha v^\beta v'^\m]$.
$\taud$ and $\taut$ are not normalised at zero recoil; however, any scale dependence vanishes: 
$\tau_{\frac{1}{2}, \frac{3}{2}}(\mu, 1)\equiv \tau_{\frac{1}{2}, \frac{3}{2}}(1)$.
A quenched lattice study obtained $\tau_{\frac{1}{2}}(1) \lesssim \tau_{\frac{3}{2}}(1)$, even if the analysis
was based on quite short plateaus
of the $J^P=2^+$ state effective mass and of 
$\tau_{\frac{1}{2},\frac{3}{2}}(1)$ data got from ratios of 3-pt and 2-pt correlation functions \cite{BecirevicTA}.
A similar computation was then led with $N_f=2$ dynamical quarks, using a set of ETMC gauge ensembles,
with acceptable signals for effective masses and $\tau_{1/2, 3/2}(1)$.
After a smooth extrapolation to the chiral limit, 
the authors found again
that $\tau_{1/2}(1)$ seems significantly smaller than $\tau_{3/2}(1)$ \cite{BlossierVY}: lattice results point 
in the same direction as quark models \cite{MorenasYQ}, \cite{EbertGA} and Operator Production Expansion based sum rules \cite{LeYaouancCJ}, \cite{UraltsevRA}.

\subsection{Towards realistic $b$ and $c$ quark masses}

\begin{figure}[t]
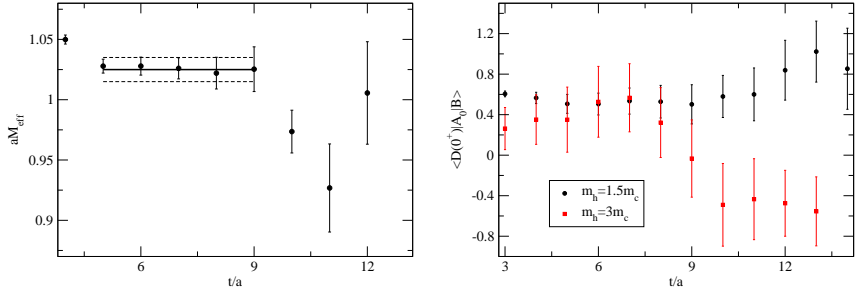

\centering
\begin{tabular}{cc}
\includegraphics[height=1.5in]{plots/effmass0pTmQCD.eps}
&
\includegraphics[height=1.5in]{plots/FFBDscalanabis.eps}
\end{tabular}
\caption{Effective mass of $D^*_0$ (left panel) and form factor ${\cal F}^0_A(1)$ at two $b$ quark masses (right panel).
\label{fig:effmasstwist}}
\end{figure}

More recently a direct computation in QCD has been tried \cite{AtouiKSA}. The starting point is the definition of a set
of form factors:
\begin{eqnarray}
\nonumber
\elematrice{D^*_0}{A^\mu}{B}&=&\tilde{u}^+\, (p_B + p_D)^\mu+
\tilde{u}^-\, (p_B - p_D)^\mu,\\
\nonumber
\elematrice{D^*_2(\epsilon^{(\lambda)})}{V^\mu}{B}&=&
i\tilde{h}\,\epsilon^{\mu\nu\rho\sigma}
\epsilon^{(\lambda)*}_{\nu\alpha}p^\alpha_B (p_B + p_D)_\rho 
(p_B -p_D)_\sigma,\\
\nonumber
\elematrice{D^*_2(\epsilon^{(\lambda)})}{A^\mu}{B}&=&
\tilde{k}\, \epsilon^{(\lambda)\, \mu \nu\,*} p_{B\,\nu}
+\epsilon^{(\lambda)\,*}_{\alpha \beta}p^\alpha_B p^\beta_B 
[\tilde{b}^+\, (p_B + p_D)^\mu + \tilde{b}^-\,(p_B - p_D)^\nu],
\end{eqnarray}
with $V_\mu= \bar{c}\gamma_\mu b$ and $A_\mu=\bar{c}\gamma_\mu \gamma^5 b$.
Choosing the kinematical configuration $\vec{p}_D = \vec{0}$, 
$\vec{p}_B = (\theta,\theta,\theta)$ and defining the tensors of polarisation
accordingly, it has been shown that the leading form factors that contribute to the widths are
\begin{equation}
\nonumber
\tilde{k}=-\frac{\sqrt{6}}{\theta}{\cal F}^{(0)\,1}_A=
-\frac{\sqrt{6}}{\theta}{\cal F}^{(0)\,2}_A
=\frac{\sqrt{6}}{2\theta}{\cal F}^{(0)\,3}_A,
\end{equation}
\begin{equation}
\nonumber
\tilde{k}=\frac{1}{\theta}\left[{\cal F}^{(+2)\,1}_A
+{\cal F}^{(-2)\,1}_A\right]=
-\frac{1}{\theta}\left[{\cal F}^{(+2)\,2}_A+{\cal F}^{(-2)\,2}_A\right],
\end{equation}
\begin{eqnarray}
\nonumber
\tilde u^+ &=& -\,\dfrac{1}{2\,m_{D^*_0}}\left[\dfrac{E_B - m_{D^*_0}}
{3\theta}
({\cal F}^1_A+{\cal F}^2_A+{\cal F}^3_A) - {\cal F}^0_A\right],
\end{eqnarray}
where ${\cal F}^{(\lambda)\,\mu}_A \equiv 
\elematrice{D^*_2(\epsilon^{(\lambda)})}{A^\mu}{B}$ and ${\cal F}^{\mu}_A \equiv 
\elematrice{D^*_0}{A^\mu}{B}$.
The preliminary study was performed using $N_f=2$ ETMC ensembles: the charm quark
was tuned at the physical point, while several "light" $b$ quarks were simulated to 
extrapolate to $m_b$; cut-off effects were investigated on 2 lattice spacings, a third one 
will finally be considered. Twisted boundary conditions are required to give a momentum to the $B$ meson
in 2-pt and 3-pt correlators.
In the twisted-mass formalism it is difficult to isolate the signal for $D^*_0$ because of the mixing with 
$D$ state due to a breaking parity cut-off effect: solving a generalized eigenvalue
problem is beneficial. as shown in the left panel of Figure \ref{fig:effmasstwist}. 
Isolating the signal for $D^*_2$ is difficult because of the noise, despite averaging 
over different interpolating fields that belong to the same representation (E or T2) of the $O_h$ cubic group.
At zero recoil, it seems possible to isolate the signal for ${\cal F}^0_A$ but it deteriotates if the $b$ quark mass 
gets closer to $m_b$, as shown in the right panel of Figure \ref{fig:effmasstwist}. Concerning the decay of $D^*_2$, it is known that ${\cal F}^{(\lambda)\,\mu}_A(1) = 0$: one needs to inject large momenta, where the data are also noisy.

\section{$B_{(s)} \to D_{(s)} \pi$: a more favorable situation?}

A comparison between theory and experiment non leptonic $B \to D$ decays is made in the right panel of Table \ref{tabSL}.
Though a (not so conclusive) experimental disagreement in ${\cal B}(B_d \to D^*_0 \pi)$ between Belle and BaBar, 
and the fact that theoretical predictions are based on the factorisation approximation, that works well for the so called Class I decays, we globally observe a much better agreement between theory and experiment for $B_d \to D^*_0 \pi$ than for $B_d \to D^*_0 l \nu$.

\subsection{Largeness of $B \to D' l \nu$ checked on $B \to D' \pi$}

It was proposed in \cite{BecirevicMP} to check the hypothesis of a large branching ratio ${\cal B}(B \to D' l \nu)$ by studying non leptonic decays. By examining the Class I process $\bar{B}^0 \to D'^+ \pi^-$, one has in the factorisation approximation
\begin{equation}\nonumber
\frac{{\cal B}(\bar{B}^0 \to D'^+ \pi^-)}
{{\cal B}(\bar{B}^0 \to D^+ \pi^-)}=
\left(\frac{m^2_B - m^2_{D'}}{m^2_B - m^2_{D}}\right)^2 
\left[\frac{\lambda(m_B,m_{D'},m_\pi)}{\lambda(m_B,m_{D},m_\pi)}\right]^{1/2}
\left|\frac{f^{B \to D'}_+(0)}{f^{B \to D}_+(0)}\right|^2,
\end{equation}
where $\lambda(x,y,z)=[x^2 - (y + z)^2][x^2 - (y - z)^2]$ and 
$f^{B \to D(')}_+(m^2_\pi) \sim f^{B \to D(')}_+(0)$. With $V_{cb}f^{B \to D}_+(0)=0.02642(8)$ from Babar \cite{vcbfpbabar} and  $\left|V_{cb}\right|^{\rm incl} = 0.0411(16)$, we obtain $f^{B \to D}_+(0) = 0.64(2)$. Next, 
with $m_{D'}=2.54$ GeV, we get $\frac{{\cal B}(\bar{B}^0 \to D'^+ \pi^-)}
{{\cal B}(\bar{B}^0 \to D^+ \pi^-)}= (1.65 \pm 0.13) \times
\left|f^{B \to D'}_+(0)\right|^2$. Finally, with ${\cal B}(\bar{B}^0 \to D^+ \pi^-)=0.268(13)\%$, we have
\begin{equation}\nonumber
{\cal B}(\bar{B}^0 \to D'^+ \pi^-) = \left|f^{B \to D'}_+(0)\right|^2 \times
(4.7 \pm 0.4) \times 10^{-3}.
\end{equation}
Letting vary the $f_+^{B \to D'}(0)$ form factor in the conservative range [0.1, 0.4], according to the existing theoretical estimates \cite{bernlochner}, \cite{fBDpanel}, we conclude that ${\cal B}(\bar{B}^0 \to D'^+ \pi^-)^{\rm th} \sim 10^{-4}$: the measurement can be performed with the B factories samples and at LHCb.
Having a look to the Class III process $B^- \to D'^0 \pi^-$, the factorised amplitude reads:
\begin{equation}
\nonumber
A^{III}_{\rm fact} = -i \frac{G_F}{\sqrt{2}}
V_{cb}V^*_{ud} \left[a_1f_\pi
[m^2_B - m^2_{D'}]f^{B \to D'}(m^2_\pi)
+a_2f_{D'}[m^2_B - m^2_\pi]f^{B \to \pi}(m^2_{D'})\right].
\end{equation}
When the corresponding branching ratio is normalised by the Class I counterpart, we find
\begin{equation}\nonumber
\frac{{\cal B}( B^-\to D^{\prime 0}\pi^-)}{{\cal B}(\bar{B}^0\to  D^{\prime +}
\pi^-)} =\frac{\tau_{B^-}}{\tau_{\bar B^0}}\left[ 1 + 
{a_2\over a_1}\times 
{m_B^2-m_\pi^2\over m_B^2-m_{D^\prime}^2} 
 \times  
{f_0^{B\to \pi}(m_{D^\prime}^2)\over f_+^{B \to D'}(0)}  
{f_{D^\prime}\over f_D} \ {f_D\over f_\pi}\right]^2.
\end{equation}
The ratio of Wilson coefficients $a_2/a_1$ is extracted from $\frac{{\cal B}( B^-\to D^0\pi^-)}
{{\cal B}(\bar{B}^0\to  D^+ \pi^-)}$, known experimentally \cite{vckmfit}, and it remains the computation
on the lattice of the ratios of decay constants $\frac{f_{D'}}{f_D}$ and $\frac{f_D}{f_\pi}$.
Combining ETMC data at different $a$ and $m_{\rm sea}$ in a common fit 
we get
\begin{eqnarray}
\nonumber
{m_{D_s^\prime}\over m_{D_s}} = 1.53(7), \quad {f_{D_s^\prime}\over f_{D_s}} = 0.59(11),\\
\nonumber
{m_{D^\prime}\over m_{D}} = 1.55(9), {f_{D^\prime}\over f_{D}} =0.57(16).
\end{eqnarray}
The experimental result is $(m_{D'}/m_{D})^{\rm exp}=1.36$, $2\sigma$ smaller than our value. For the moment
that discrepancy remains unexplained despite several checks described in \cite{BecirevicMP}. 
With $a_{2}/a_{1}=0.368$, $\tau_{\bar B^0}/\tau_{B^-}=1.079(7)$, $f_+^{B\to D}(0)= 0.64(2)$ and $f_0^{B\to \pi}(m_D^2)= 0.29(4)$ \cite{blazenka}, we obtain
\begin{equation}\nonumber
{{\cal B}( B^-\to D^{\prime 0}\pi^-)\over {\cal B}(\bar{B}^0\to  D^{\prime +}
\pi^-)} ={\displaystyle{\tau_{B^-}\over \tau_{\bar B^0}} \left[ 1 
+ \displaystyle{0.14(4)\over f_+^{B \to D'}(0)} \right]^2}, \quad
{{\cal B}(\bar{B}^0\to D^{\prime +}\pi^-)
\over {\cal B}(\bar{B}^0\to D^{+}\pi^-)} =
(1.24\pm 0.21)\times |f_+^{B \to D'}(0)|^2.
\end{equation}
Using the experimental value $\frac{m_{D^\prime}}{m_{D}}=1.36$, we get $\frac{{\cal B}(\bar{B}^0 \to D'^+ \pi^-)}
{{\cal B}(\bar{B}^0 \to D^+ \pi^-)}=(1.65 \pm 0.13) \times
\left| f^{B \to D'}_+(0)\right |^2$: the dependence on $m_{D'}$ of that ratio is actually small. 
Fixing $f^{B \to D'}_{+}(0)=0.4$ and taking $(m_{D'}/m_{D})^{\rm exp}$ we have also
\begin{equation}
\nonumber
\frac{{\cal B}( \bar{B}^0\to  D^{\prime +}\pi^-)}{{\cal B}(\bar{B}^0\to  D_2^{*+}\pi^-)}
=1.6(3), \quad\quad
\frac{{\cal B}( B^-\to D^{\prime 0}\pi^-)}{{\cal B}(B^- \to  D_2^{*0}\pi^-)}
=1.4(3).
\end{equation}
It means that if $f_+^{B \to D'}$ is large, as claimed by many authors, the measurement of
${\cal B}(B \to D' \pi)$ should be as feasible as ${\cal B}(B \to D^{*}_{2} \pi)$.

\subsection{$B_s \to D^{**}_s \pi$}

The situation of the $D_s$ spectrum is peculiar: indeed, $D^*_{s0}(2317)$ and $D^*_{s1}(2460)$ are below the $DK$ and $D^* K$ thresholds. The main consequence is that they are narrow states. Thus it is very advantageous to examine
them because there is no experimental issue from their broadness. It has been proposed to study hadronic decays $B_s \to D^{*+}_{s0}(2317) \pi^-$ and $B_s \to D^{*+}_{s1}(2460) \pi^-$ \cite{BecirevicTE}.
At the moment, only upper limits on ${\cal B}(D^{*+}_{s0} \to ...)$ are available: 
${\cal B}(D^{*+}_{s0} \to D^+_s \gamma,\; D^{*+}_{s0} \to D^{*+}_s \gamma \gamma) < 0.2\%$.
In phenomenological analyses, the range ${\cal B}(D^{*+}_{s0} \to D^+_s \pi^0)=(97 \pm 3)\%$ is taken.
There are more data concerning the decay of $D^*_{s1}$, that we collect in Table \ref{tab:ds1}.
\begin{table}
\begin{center}
\begin{tabular}{|l|}
\hline
${\cal B}(D^{*+}_{s1} \to D^{*+}_s \pi^0)=(48 \pm 11)\%$\\ 
${\cal B}(D^{*+}_{s1} \to D^{+}_s \gamma)=(18 \pm 4)\%$\\
${\cal B}(D^{*+}_{s1} \to D^{+}_s \pi^+ \pi^-)=(4.3 \pm 1.3)\%$\\
${\cal B}(D^{*+}_{s1} \to D^{*+}_{s0} \gamma)=(3.7^{+5.0}_{-2.4})\%$\\
\hline
\end{tabular}
\end{center}
\caption{Branching ratios of non leptonic $D^*_{s1}$ decays. \label{tab:ds1}}
\end{table}
According to \cite{BecirevicTE}, at LHCb, one measures the cascade $B_s \to D^{*-}_{s0} \pi^+$, $D^{*-}_{s0} \to D^-_s \pi^0$, $D^-_s \to K^+ K^- \pi^-$; the 4-momentum of the non detected $\pi^0$ is extracted from the $B_s$ flight direction and the known $m_{B_s}$ and $m_{\pi^0}$.  The narrow peak in the $D^-_{s0} \pi^0$ mass distribution can be observed, depending on the accuracy of tracking capabilities. Neglecting SU(3) breaking effects, with ${\cal B} (B_s \to D_s^+ \pi^-) = (2.95 \pm 0.28) \times 10^{-3}$ and ${\cal B}(B_s \to D^{*-}_{s0} \pi^+)=(1 \pm 0.5) \times 10^{-4}$, the number of expected events with 1 ${\rm fb}^{-1}$ of integrated luminosity is
\begin{equation}
N(B_s \to D^{*-}_{s0} \pi^+) = 600 \times (1 \pm 0.5) 
\times {\cal B}(D^{*-}_{s0} \to D^-_s \pi^0) \times \epsilon_{\pi^0}:\; \sim 100.
\end{equation}



\begin{thebibliography}{99}

\bibitem{vckmfit} 
J.~Beringer {\it et al.}  [Particle Data Group Collaboration],
Phys.\ Rev.\ D {\bf 86}, 010001 (2012).

\bibitem{babard}
  P.~del Amo Sanchez {\it et al.}  [BABAR Collaboration],
  Phys.\ Rev.\ D {\bf 82} (2010) 111101.

\bibitem{quarkmodels}
F.~E.~Close and E.~S.~Swanson,
  Phys.\ Rev.\ D {\bf 72} (2005) 094004;
  Z.~-F.~Sun, J.~-S.~Yu, X.~Liu and T.~Matsuki,
  Phys.\ Rev.\ D {\bf 82} (2010) 111501.

\bibitem{bernlochner}
  F.~U.~Bernlochner, Z.~Ligeti and S.~Turczyk,
  Phys.\ Rev.\ D {\bf 85} (2012) 094033.

\bibitem{puzzle}
 A.~Le Yaouanc \emph{et al},
  Phys.\ Rev.\ D {\bf 56} (1997) 5668;
  A.~K.~Leibovich \emph{et al},
  Phys.\ Rev.\ D {\bf 57} (1998) 308;
D.~Becirevic \emph{et al},
  Phys.\ Rev.\ D {\bf 87}, no. 5, 054007 (2013);
  I.~I.~Bigi \emph{et al},
  Eur.\ Phys.\ J.\ C {\bf 52}, 975 (2007).

\bibitem{BecirevicTA}
  D.~Becirevic, B.~Blossier, P.~Boucaud, G.~Herdoiza, J.~P.~Leroy, A.~Le Yaouanc, V.~Morenas and O.~Pene,
Phys.\ Lett.\ B {\bf 609}, 298 (2005).
[hep-lat/0406031].

\bibitem{BlossierVY}
  B.~Blossier {\it et al.}  [European Twisted Mass Collaboration],
JHEP {\bf 0906}, 022 (2009).
[arXiv:0903.2298 [hep-lat]].

\bibitem{MorenasYQ}
  V.~Morenas, A.~Le Yaouanc, L.~Oliver, O.~Pene and J.~C.~Raynal,
Phys.\ Lett.\ B {\bf 386}, 315 (1996).
[hep-ph/9605206].

\bibitem{EbertGA}
  D.~Ebert, R.~N.~Faustov and V.~O.~Galkin,
Phys.\ Rev.\ D {\bf 61}, 014016 (2000).
[hep-ph/9906415].

\bibitem{LeYaouancCJ}
  A.~Le Yaouanc, D.~Melikhov, V.~Morenas, L.~Oliver, O.~Pene and J.~C.~Raynal,
Phys.\ Lett.\ B {\bf 480}, 119 (2000).
[hep-ph/0003087].

\bibitem{UraltsevRA}
  N.~Uraltsev,
[hep-ph/0409125].

\bibitem{AtouiKSA}
  M.~Atoui, B.~Blossier, V.~Morénas, O.~Pène and K.~Petrov,
[arXiv:1312.2914 [hep-lat]].

\bibitem{BecirevicMP}
  D.~Becirevic \emph{et al},
Nucl.\ Phys.\ B {\bf 872}, 313 (2013).

\bibitem{vcbfpbabar}
  B.~Aubert {\it et al.}  [BABAR Collaboration],
  Phys.\ Rev.\ Lett.\  {\bf 104} (2010) 011802.

\bibitem{fBDpanel}
  J.~Hein {\it et al.}  [UKQCD Collaboration],
  Nucl.\ Phys.\ Proc.\ Suppl.\  {\bf 83} (2000) 298;
  D.~Ebert, R.~N.~Faustov and V.~O.~Galkin,
  Phys.\ Rev.\ D {\bf 62} (2000) 014032;
R.~N.~Faustov and V.~O.~Galkin,
  Phys.\ Rev.\ D {\bf 87}, 034033 (2013);
 Z.~-H.~Wang \emph{et al},
  J.\ Phys.\ G {\bf 39} (2012) 085006.
  
\bibitem{blazenka}
  G.~Duplancic \emph{et al},
  JHEP {\bf 0804} (2008) 014;
P.~Ball and R.~Zwicky,
  Phys.\ Rev.\ D {\bf 71} (2005) 014015.

\bibitem{BecirevicTE}
  D.~Becirevic, A.~Le Yaouanc, L.~Oliver, J.~C.~Raynal, P.~Roudeau and J.~Serrano,
Phys.\ Rev.\ D {\bf 87}, no. 5, 054007 (2013).
[arXiv:1206.5869 [hep-ph]].

\end{thebibliography}
\end{document}